\documentclass[a4paper, oneside, twocolumn, notitlepage, 10pt]{extarticle_ecoc}
\usepackage{ecoc}

\begin{document}
\selectlanguage{english}    


\title{End-to-End Learning of Pulse-Shaper and Receiver Filter in the Presence of Strong Intersymbol Interference}


\author{
    Søren Føns Nielsen\textsuperscript{(1)}, Francesco Da Ros\textsuperscript{(2)},
    Mikkel N. Schmidt\textsuperscript{(1)}, Darko Zibar\textsuperscript{(2)}
}

\maketitle                  


\begin{strip}
    \begin{author_descr}

        \textsuperscript{(1)} DTU Compute, Technical University of Denmark
        \textcolor{blue}{\uline{sfvn@dtu.dk}}

        \textsuperscript{(2)} DTU Electro, Technical University of Denmark

    \end{author_descr}
\end{strip}

\renewcommand\footnotemark{}
\renewcommand\footnoterule{}


\begin{strip}
    \begin{ecoc_abstract}
       We numerically demonstrate that joint optimization of FIR based pulse-shaper and receiver filter results in an improved system performance, and shorter filter lengths (lower complexity), for 4-PAM 100 GBd IM/DD systems. 
    \end{ecoc_abstract}
\end{strip}


\section{Introduction}
A combination of higher order PAM signaling and baud rates reaching 100 GBd or beyond will be needed for the next-generation of data center interconnects to reach 800 Gb/s and 1.6 Tb/s \cite{Che:24}. This will require ultra-high bandwidths of optical and electrical components, which may be challenging to realize in practice. It is therefore expected that the next generation data center fiber-optic communication systems will need to deal with strong inter-symbol-interference (ISI), while at the same time keeping the complexity, and thereby power consumption, down \cite{Che:24,Yan:24}. 

The bandwidth limitations of the transmitter- and the receiver-front-end (digital-to-analog-converter, optical modulator, photodiodes and analog-to-digital-converter), and chromatic dispersion of the fibre channel, will induce ISI. The mitigation can be performed by employing transmitter- and receiver-side compensation i.e.~pre-distortion and equalization\cite{Liang:24,Yan:24,Che:24,Zhong18}. To achieve the zero ISI condition, the total transfer function including: pre-distortion, pulse-shaper, transmitter front-end, channel, receiver front-end, receiver filter and equalizer, must fulfill the Nyquist criteria \cite{Proakis}. 

The optimum performance (zero ISI and maximization of the received signal signal-to-noise ratio), is then obtained by having receiver a $h_{rx}(t)$ that is matched to the \textit{combined} impulse response, $h_{comb}(t)$, of the transmitter, channel and receiver-front-end, $h_{rx}(t)=h_{comb}(-t)$. Typically, pre-distortion and equalization are optimized separately, which may lead to sub-optimum system performance \cite{Zhong18}. Additionally, it would be relevant to investigate if joint optimization allows for a complexity reduction, i.e. decreased number of taps. 

In this paper, we employ end-to-end learning and jointly learn pulse-shaper and receiver filter that would result in optimum performance (zero ISI and maximization of SNR). To keep the complexity low, the pulse-shaper and receiver filter are implemented as linear FIR filters.  

The proposed approach requires computation of gradients through the transmitter, channel and receiver, that is difficult and cumbersome to perform. However, by implementing the entire system in a machine learning framework, such as \texttt{pytorch}, gradient computation is easily achieved by automatic differentiation.

We show numerically, that joint optimization of the pulse-shaper and receiver filter significantly improves the system performance and results in a lower total number of filter taps for back-to-back and 2 km transmission for 100 GBd 4-PAM IM/DD links.

\section{Simulation framework}

\begin{figure*}[!t]
    \centering
    \includegraphics[width=1.0\linewidth]{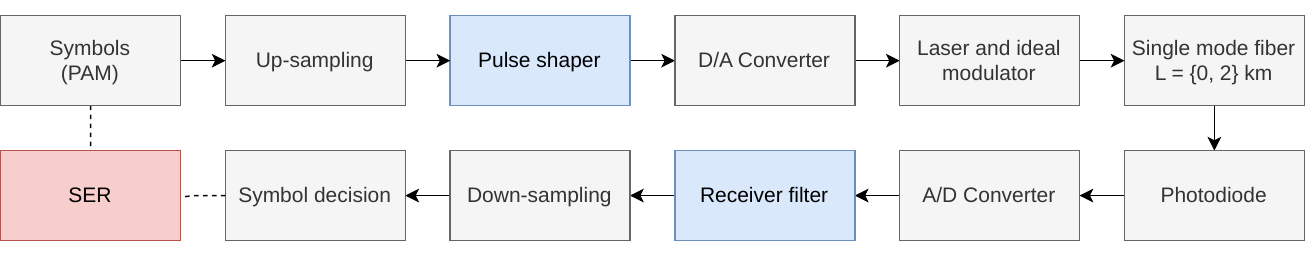}
    \caption{Blockdiagram of the simulation framework during evaluation. The blue boxes indicate the blocks that are learnable during the training phase.}
    \label{fig:block-diagram}
\end{figure*}

The simulation setup is shown in Fig.~\ref{fig:block-diagram}. First, a sequence of 4-PAM symbols is generated and up-sampled to 4 samples per symbol. It is then passed through a pulse-shaper, implemented as an FIR filter, with learnable weights. The pulse-shaper filter is initialized as the root-raised cosine (RRC) filter with rolloff $\alpha=0.01$. The filter length is varied in the numerical simulations. The baud rate of the information carrying signal after the pulse-shaper is 100 GBd. The signal is then passed to a digital-to-analog converter (DAC), modeled by a 5th order Bessel filter with a 3 dB cutoff of 45 GHz. We use an ideal (linear) optical modulator to convert the electrical signal to the optical domain. The continuous wave laser at the input to the modulator has power denoted by $P_{in}$ and the wavelength $\lambda = 1270$ nm. The channel is an optical fibre with zero-dispersion at wavelength $\lambda_0 = 1310$ nm.

After photodetection, the signal is passed through an analog-to-digital (ADC) converter with the same model and parameters as the DAC.
The signal after the ADC is then passed through the receiver filter implemented as a linear FIR filter with learnable parameters and the same number of taps as the pulse-shaper. At the receiver filter output, the signal is down-sampled to 1 sample per symbol. 

\begin{figure*}[t]
    \centering
    \includegraphics[width=\linewidth]{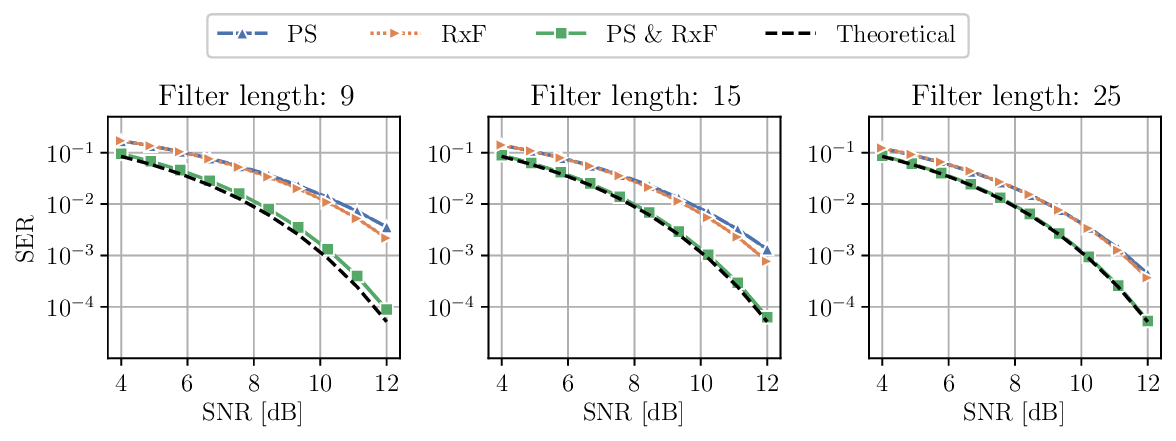}
    \caption{Symbol error rate as a function of SNR in the B2B (AWGN) scenario. We perform optimization of: 1) pulse-shaper (PS), 2) receiver filter (RxF) and 3) joint pulse-shaper and receiver filter (PS \& RxF).}
    \label{fig:awgn_ser_vs_esn0}
\end{figure*}

The entire system is implemented in \texttt{pytorch} to allow for the use of automatic differentiation when taking the gradients through the ADC, optical channel and DAC in order to optimize the FIR based pulse-shaper and the receiver filter. The objective is to learn a set of pulse-shaper and receiver filter weights that would result in zero ISI. The total frequency characteristic of the system would in that case adhere to Nyquist's zero ISI condition\cite{Proakis} (cf. equation~\eqref{eq:zero_isi}).  
During training, the output of the down-sampler is compared to the transmitted symbol sequence using the mean squared error (MSE). The gradient of the MSE is then propagated through the system in order to adapt the filter weights.  
We use the Adam optimizer\cite{kingma_adam_2015} with the OneCLR learning rate policy\cite{smith_super-convergence_2019} setting the maximum learning rate to 10 times the starting learning rate, $lr_0$. At each gradient update, we use gradient norm clipping\cite{pascanu_difficulty_2013} and normalize the filters to have unit L2-norm.
In all optimization runs, we use a batch size of $1000$ symbols and we exhaustively screen the following starting learning rates $lr_0 = [5\cdot 10^{-3}, 1\cdot 10^{-3}, 5\cdot 10^{-4}, 1\cdot 10^{-4}, 5\cdot 10^{-5}]$. 

During evaluation, a new sequence of symbols is generated and passed through the system with both filters fixed to their value at the last step of training process. Symbol decisions are made using symbol-based maximum likelihood decisions.
We use $2.5\cdot 10^6$ symbols for training and $1.0 \cdot 10^6$ symbols for evaluation. Each simulation is repeated 5 times with different random seeds and in the following symbol error rate (SER) results are reported as an average over seeds.

\begin{figure}
    \centering
    \includegraphics[width=0.95\linewidth]{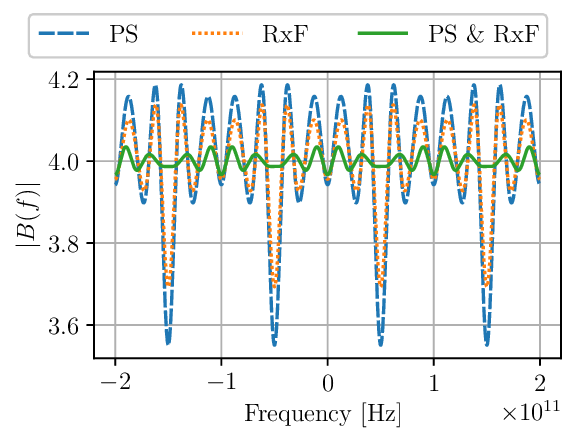}
    \caption{Nyquist's zero ISI condition in the frequency domain calculated on the system response of the B2B scenario. For details on how $|B(f)|$ was calculated please refer to equation \eqref{eq:zero_isi}. Filter lengths were set to 25.}
    \label{fig:sys_isi}
\end{figure}

\begin{figure*}[t]
    \centering
    \includegraphics[width=0.32\linewidth]{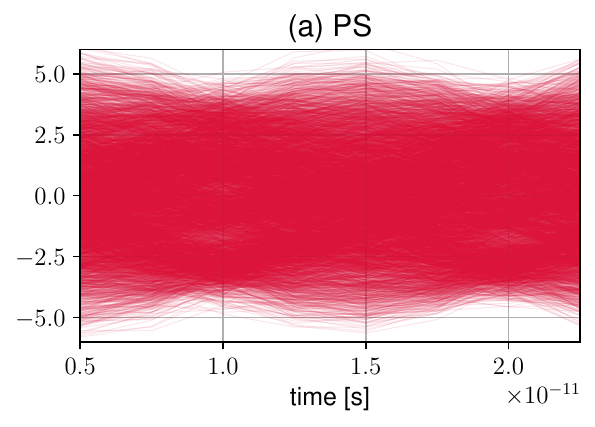}
    \includegraphics[width=0.32\linewidth]{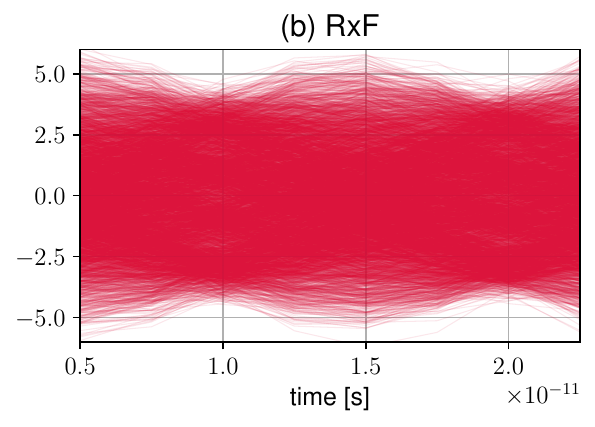}
    \includegraphics[width=0.32\linewidth]{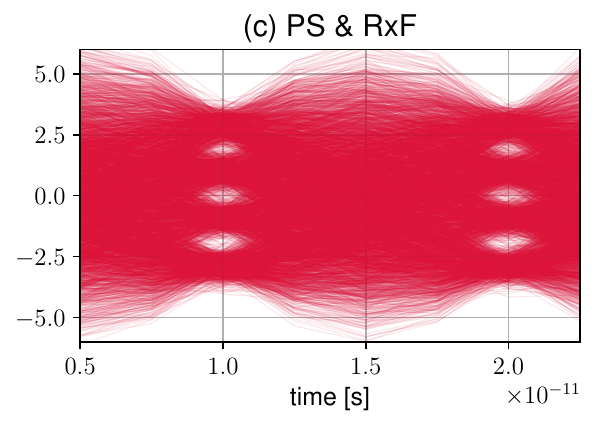}
    \caption{Eyediagrams of the signal at the output of the receiver filter for $P_{rec} \approx -6.6\; \mathrm{dBm}$ for the 2 km SMF scenario. (a) Pulse-shaper optimization only, (b) receiver filter optimization only, (c) joint optimization of pulse-shaper and receiver filter.}
    \label{fig:eyediagrams}
\end{figure*}

\begin{figure*}
    \centering
    \includegraphics[width=\linewidth]{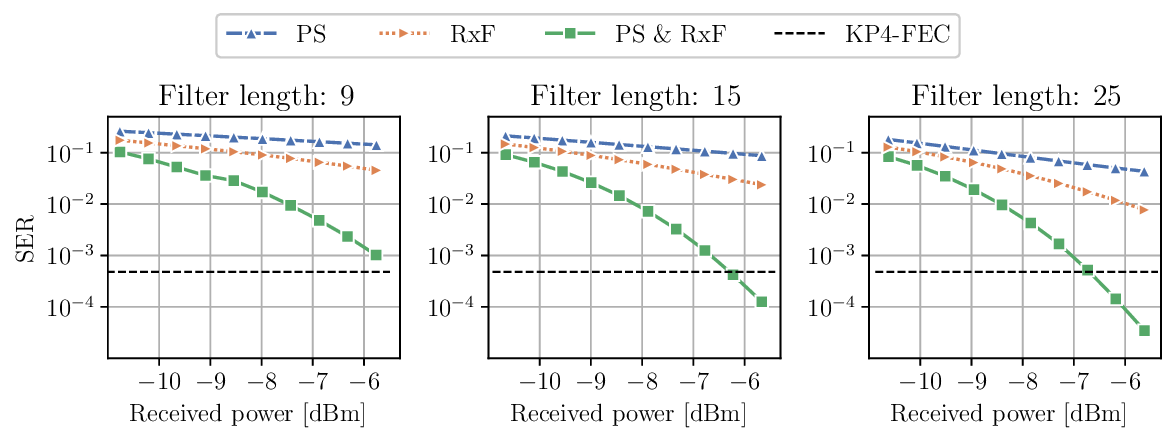}
    \caption{Symbol error rate as a function of received signal power in direction detection scenario (fiber length of 2 km). We optimize: 1) pulse-shaper (PS), 2) receiver filter (RxF) 3) joint pulse-shaper and receiver filter (PS \& RxF).}
    \label{fig:ser_vs_power}
\end{figure*}

\section{Results}
To begin with, we consider a back-to-back (B2B) IM/DD channel, equivalent to an additive white Gaussian noise (AWGN) channel with bandwidth limitation. The system is optimized in three variants to study the effects of joint optimization: 1) pulse-shaper is optimized and the receiver filter is held fixed as an RRC filter, (\textbf{PS}), 2) receiver filter is optimized and the pulse-shaper is held fixed as an RRC filter, (\textbf{RxF}), 3) pulse-shaper and receiver filter are jointly optimized, (\textbf{PS \& RxF}). During optimization, the SNR is fixed to 12 dB and during evaluation we vary the thermal noise. The results are shown in Fig.~\ref{fig:awgn_ser_vs_esn0}. A general observation is that for joint optimization of pulse-shaper and receiver filter zero ISI can be achieved, i.e. the SER curve matches theory, while this is not the case for only optimizing the pulse-shaper or receiver filter. Moreover, joint optimization requires fewer filter taps, to reach a certain performance, which may lead to less complexity and thereby lower power consumption. For instance, a power penalty of $0.3$ dB at SER of $10^{-4}$ is achieved with only 9 taps, when performing joint optimization. While even with 25 taps, such a performance cannot be achieved if only pulse-shaper or receiver filter are optimized. One explanation for this, could be that when fixing one of the filters, the other does not have enough degrees of freedom to invert the system response.
To illustrate what the optimization methods are learning, we plot the Nyquist zero ISI criterion for the total system response (including the DAC and ADC) for each of the methods in Figure~\ref{fig:sys_isi}. Let $h(t)$ denote the total time-domain transfer function of pulse-shaper, DAC, ADC and receiver filter. Given the baud rate $\frac{1}{T}$ and Fourier transform of the system, $H(f)$, then the function of interest\cite{Proakis} is
\begin{equation}
    B(f) = \sum_{m=-\infty}^\infty H\left(f + \frac{m}{T}\right).\label{eq:zero_isi}
\end{equation}
The zero ISI criterion states that $B(f)$ should be constant for all frequencies. Out of the three methods, the joint optimization method has the flattest response. This stands in contrast to the pulse-shaper and receiver filter optimizations that lead to significant deviations from the desired constant value.

Next, we consider 2 km of SMF transmission. The reason why we only consider such short transmission distance is because at 100 GBd the signal is severely limited by the dispersion. Additionally, we only consider FIR filters that have limited compensation capability in case of direct detection. In this simulation, the systems are trained and evaluated at the same launch powers. Fig.~\ref{fig:eyediagrams}(a)-(c) shows eyediagrams after the receiver filter. The figures clearly show that for joint optimization of pulse-shaper and receiver filter eye opening is observed, while this is not the case if only pulse-shaper or receiver filter are optimized. Finally, in Fig.~\ref{fig:ser_vs_power}, we plot the SER as a function of received signal power at the detector for different pulse-shaper and receiver filter lengths. In this simulation, we vary the laser power at the modulator, $P_{in}$. It is clearly observed that joint optimization significantly improves system performance. Using FIR filters with 15 taps, we can achieve an SER below the KP4 forward error correction (FEC) threshold.

\section{Conclusions}
We have demonstrated the effectiveness of jointly optimizing FIR based pulse-shaper and receiver filters in a simulated PAM4 modulated 100 GBd direct detection system. Significant performance gains, in terms of SER and number of filter taps, are obtained compared to pulse-shaper and receiver filter optimization, only. The key enabler for the joint optimization is the automatic differentiation which allows for effective computation of gradients though the ADC, optical fibre and DAC.

\clearpage
\section{Acknowledgements}
This research was supported by research grants from VILLUM FONDEN: VI-POPCOM (VIL54486), MARBLE (VIL40555), and YIP OPTIC-AI (VIL29334).

\printbibliography

\vspace{-4mm}

\end{document}